\documentclass[a4paper]{article}
\usepackage[margin=25mm]{geometry}
\usepackage{amsmath}
\usepackage{amsfonts}
\usepackage{amssymb}
\usepackage{graphicx}
\usepackage{verbatim}
\usepackage[linesnumbered,ruled,vlined]{algorithm2e}
\usepackage{colortbl}
\usepackage[table]{xcolor}
\usepackage{booktabs}
\usepackage{soul}
\usepackage{graphicx}
\usepackage{epsf,epsfig}
\usepackage{arydshln} 
\usepackage{colortbl}
\usepackage{multirow}
\usepackage{rotating}
\usepackage[table]{xcolor}  % For table colors
\usepackage{multirow}       % For multi-row cells
\usepackage{graphicx}       % For text rotation
\usepackage{authblk}
\usepackage{mathrsfs}
\providecommand{\keywords}[1]
{
  \small	
  \textbf{\textit{Keywords---}} #1
}

\title{GPU Accelerated Transducer-Field Calculation using the Traditional Born Series Formulation for Realistic Media}
\author[1]{Ujjal Mandal}
\author[2]{Jagpreet Singh}
\author[3]{Ben T Cox}
\author[1]{Ratan K Saha}
\affil[1]{Department of Applied Sciences, Indian Institute of Information Technology Allahabad, Jhalwa, Prayagraj, 211015, Uttar Pradesh, India}
\affil[2]{Department of Computer Science and Engineering, Indian Institute of Technology Ropar, Street 29, Ropar, 140001, Punjab, India}
\affil[3]{Department of Medical Physics and Biomedical Engineering, University College London, Gower Street, London, WC1E 6BT, UK}
\date{} % Comment this line to show today's date

\begin{document}
\maketitle
% \linenumbers
\begin{abstract}
This study numerically solves inhomogeneous Helmholtz equations modeling acoustic wave propagation in homogeneous and lossless, absorbing and dispersive, inhomogeneous and nonlinear media. The traditional Born series (TBS) method has been employed to solve such equations. The full wave solution in this methodology is expressed as an infinite sum of the solution of the unperturbed equation weighted by increasing power of the potential. Simulated pressure field patterns for a linear array of acoustic sources (a line source) estimated by the TBS procedure exhibit excellent agreement with that of a standard time domain approach (k-Wave toolbox). The TBS scheme though iterative but is a very fast method. For example, GPU enabled CUDA C code implementing the TBS procedure takes 5 s to calculate the pressure field for the homogeneous and lossless medium whereas nearly 500 s is taken by the later module. The execution time for the corresponding CPU code is about 20 s. The findings of this study demonstrate the effectiveness of the TBS method for solving inhomogeneous Helmholtz equation, while the GPU-based implementation significantly reduces the computation time. This method can be explored in practice for calculation of pressure fields generated by real transducers designed for diverse applications. 
\end{abstract}\hspace{10pt}
%
%TC:ignore
\section{Introduction}
Ultrasound (US) imaging is a very popular medical imaging modality because it is cheap, portable, noninvasive and nonionizing \cite{SWang}. However, it can provide images of deeply seated soft tissue structures (penetration depth is about 6 cm for a 5 MHz transducer). The main component of US imaging is the transducer. Accurate determination of pressure field emitted by a transducer is of profound importance for system design, image reconstruction, understanding wave-medium interactions, tissue characterization, development of beamforming algorithms, dosimetry for high intensity focused US therapy and training of ultrasonographers (who use US equipment, generate and interpret images) \cite{Jing}. Different variants of time dependent inhomogeneous wave equations are available which model acoustic wave propagation in real tissue. In general, such an equation is solved numerically to yield the pressure field distribution for a given set of initial and boundary conditions \cite{Vanmanen}. An ideal simulator should- i) accommodate various transducer geometries (single element, multi element, linear, curved and phased array transducers), ii) provide fatihful solutions for realistic (homogeneous and lossless, absorbing and dispersive, inhomogeneous and nonlinear) media, iii) handle large computational domain and perform fast calculations for long time (e.g., 6 cm distance corresponds to 200 and 400 wavelengths at 5 and 10 MHz, respectively).  

The Green's function method has been widely utilized by many groups to solve time dependent wave equations. For example, FIELD II software calculates spatial impulse response of a transducer based on the Green's function approach \cite{Jensen}. Rayleigh integral has also been carried out to estimate the field distribution of a transducer of interest \cite{Jwu}. These methods perform satisfactorily for homogeneous and lossless tissue media. Numerical solutions, using finite element and finite difference time domain, of wave equations involving fractional Laplacian operators for lossy media have been explored \cite{Chenholm}. These approaches did not consider nonlienar propagation of acoustic waves. Another well accepted simulation software is the k-Wave toolbox, where k-space pseudospectral method is implemented \cite{kwavetoolbox}. This methodology can take care off various transducer geometries, linear and nonlinear features of the propagating medium and facilitates reliable estimates of pressure field patterns. However, as expected, the computational cost of k-Wave grows significantly with domain size and simulation duration, particularly in 3D settings. The Broadband Generalized Angular Spectrum Method (BBGASM) and Iterative Nonlinear Contrast Source (INCS) method represent two powerful approaches for simulating nonlinear acoustic wave propagation in ultrasound applications. BBGASM implement in the frequency domain, enabling fast and efficient broadband simulations with accuracy for multi-harmonic fields, and is effective when GPU-accelated, offering up to 13x faster computation compared to CPU-based solvers \cite{Varray2011, Varray2013}. INCS reformulates the nonlinear Westervelt equation into an integral equation framework, offering high-fidelity modeling of complex scattering and inhomogeneous media with higher computational cost \cite{Demi2010, Verweij2009}. The computational methodology is termed as the iterative Neumann scheme. For reproducibility and broader accessibility, open source tools such as mSOUND have been developed, providing user-friendly simulation environments for both linear and nonlinear acoustic propagation in heterogeneous media \cite{msound2021}.

Another approach is to convert the time dependent wave equation into the frequency domain via the Fourier transformation technique and solve that equation to obtain the steady state field. The time independent form is called the Helmholtz equation. The Helmholtz equation arises in many fields \cite{Morse, Krebes}. For example, time independent Schr\"{o}dinger equation, describing scattering process of a plane wave by a potential, essentially reduces to the inhomogeneous Helmholtz equation. Solving an inhomogeneous Helmholtz equation is not a trivial task. The Green's function approach is generally applied and it provides an integral solution, referred to as the Lippmann–Schwinger equation in quantum mechanics \cite{Morse}. In this formulation, the field inside the scatterer has to be known a priori. This poses a challenge in practice. Max Born developed a method in 1926, while studying quantum scattering problems, to overcome this problem \cite{Born}. In this approach, the full wave solution is evaluated by iteratively applying the potential to the free space solution (i.e., in absence of the scatterer/perturbation). The solution appears as an infinite series and referred to as the traditional Born series (TBS). Also called the method of successive approximations. The first term of the series is called the first Born approximation and so on and so forth. Higher order terms facilitate increasingly accurate approximations to the total wave solution. 

The TBS algorithm works well when the sactterer is small and weak. Recently, Osnabrugge et al. developed a method which is called the convergent Born series (CBS) \cite{Osnabrugge}. The CBS scheme introduces a preconditioner in the TBS expression and thus extends the validity domain of the TBS protocol \cite{Osnabrugge, Kruger}. It provides accurate results for arbitrarily large and strong scatterers. Both the TBS and CBS schemes have been applied for photoacoustic field calculations by single cell and tissue mimicking various practical conditions \cite{Anuj_JMO, RKS1, RKS2, UM_Photoacoustics}. To the best of our knowledge, transducer pressure field calculation using the TBS method along with GPU implementation has never been conducted. It is a robust formulation, which can account for many realistic tissue properties. This is the novelty of the work.

The primary objective of this study is to develop and validate an efficient and fast numerical framework for simulating spatial distribution of pressure amplitude and phase produced by a linear array of point acoustic sources resembling an ultrasound transducer in 2D (i.e., a line source). The corresponding inhomogeneous Helmholtz equation in 2D was solved by employing the TBS method. Four types of tissue media were considered in this work- homogeneous and lossless, absorbing and dispersive, inhomogeneous (with scatterers) and nonlinear media. GPU-enabled CUDA C codes were written and executed for this purpose. The same pressure fields were also generated using the k-Wave package for comparison. The TBS scheme is a very fast method and offers accurate estimations of pressure field for realistic media. It is found that the TBS approach approximately 76 times faster than the k-Wave method. The TBS technique can be developed further to determine the pressure fields produced by typical diagnostic linear and phased array transducers (3D). 

The layout of the paper is as follows. The governing equations for pressure wave propagation through various media, the Born series technique and iterative solutions utilizing this framework for different scenarios are covered in section 2. The next section illustrates the computational methods and parameters considered in this study. Section 4 presents the numerical results. Discussion and conclusions of this work are outlined in section 5.
\section{Model equations}
\subsection{Wave equations in realistic media}
\subsubsection{Lossless medium}
The linear wave equation for a lossless homogeneous medium is given by \cite{Morse}, 
\begin{eqnarray}
\nabla^2 \Psi(\textbf{r}, t) -\frac{1}{v_f^2} \frac{\partial^2}{\partial t^2}\Psi(\textbf{r}, t)  
 = 0,
\label{acoustic_waveequn_time_dependent}
\end{eqnarray}
where $\Psi(\textbf{r}, t)$ is the space-time dependent acoustic pressure with $\textbf{r} \in \mathbb{R}^n$, $n=1, 2, 3$ and $t \in \mathbb{R}^+$; $v_f$ is the speed of sound for the propagating medium; the subscript $f$ indicates the ambient fluid medium. Note that the region of interest does not include any active acoustic source. The frequency domain equation becomes \cite{Morse},
\begin{eqnarray}
\nabla^2 \psi(\textbf{r}) + k_f^2 \psi(\textbf{r}) = 0,
\label{acoustic_waveequn_time_independent}
\end{eqnarray}
where $k_f$ is the wave number. Eq. (\ref{acoustic_waveequn_time_independent}) represents a homogeneous Helmholtz equation, which can readily be solved for simple geometries such as an infinite plane, infinite cylinder, sphere using the method of separaton of variables and applying appropriate boundary conditions. 
\subsubsection{Absorbing and dispersive medium}
It has been experimentally shown that acoustic attenuation in biological tissue is frequency dependent and can be mathematically modeled as \cite{Treeby1},
\begin{eqnarray}
\alpha=\alpha_0 \omega^{\nu},
\label{power_law_absorption}
\end{eqnarray}
where, $\alpha_0$ is the attenuation coefficient in Np (rad/s)$^{-\nu}$ m$^{-1}$, and it is a medium-dependent parameter; $\omega$ is the angular frequency and $\nu$ is the power-law exponent ($\nu \neq$ 1). Its typical range is $\nu \in [0, 2]$ for biological tissue. One corresponding time dependent wave equation for a homogeneous but absorbing and dispersive medium can be written as \cite{Treeby1},
\begin{eqnarray}
\nabla^2 \Psi(\textbf{r}, t) -\frac{1}{v_f^2} \frac{\partial^2}{\partial t^2} \Psi(\textbf{r}, t) +\tau \frac{\partial}{\partial t} (-\nabla^2)^{\nu/2}\Psi(\textbf{r}, t) + \eta (-\nabla^2)^{(\nu+1)/2}\Psi(\textbf{r}, t)
 = 0
\label{acoustic_waveequn_time_dependent_absorbing_dispersion}
\end{eqnarray}
where $\tau=-2\alpha_0v_f^{\nu-1}$, $\eta=2\alpha_0v_f^{\nu}\tan(\pi \nu/2)$ are two proportionality constants. The third and fourth terms on the left hand side, displaying derivative operators, account for power absorption and associated dispersion, respectively. Moreover, the absorption and dispersion are interconnected via the Kramers-Kronig relations. Further, these operators are together termed as the fractional Laplacian ($L_{frac}$),
\begin{eqnarray}
L_{frac}=\tau \frac{\partial}{\partial t} (-\nabla^2)^{\nu/2} + \eta (-\nabla^2)^{(\nu+1)/2}.
\label{fractional_laplacian}
\end{eqnarray}
The first term on the right hand side of Eq. (\ref{fractional_laplacian}) is called the lossy operator. A generalized version of the fractional Laplacian can be written as,
\begin{eqnarray}
L_{frac}=\tau \frac{\partial^{2+g}}{\partial t^{2+g}} (-\nabla^2)^h + \eta \frac{\partial^{2+q}}{\partial t^{2+q}} (-\nabla^2)^w.
\label{fractional_laplacian_2}
\end{eqnarray}
Table \ref{tab:lossy_derivatives} summarizes various forms of the fractional Laplacian considered by different groups.

The time-independent wave equation for the same medium can be derived as, 
\begin{eqnarray}
\nabla^2 \psi(\textbf{r}) + k_f^2 \psi(\textbf{r}) + \{-i \tau \omega (-\nabla^2)^{\nu/2} + \eta (-\nabla^2)^{(\nu+1)/2} \}\psi(\textbf{r}) = 0.
\label{acoustic_waveequn_time_independent_absorbing_dispersion}
\end{eqnarray}
The Green's function method can be applied to solve the above equation \cite{GreensFCox}.
\begin{table}[h]
\centering
\caption{Lossy and dispersive derivative operators \cite{Treeby1}.}
\label{tab:lossy_derivatives}
\begin{tabular}{|c c c c c c|}
\hline
Name & $g$ & $h$ & $q$ & $w$ & Dependence \\
\hline
Damped & -1 & 0 & 0 & 0 & $\omega^0$ \\
Blackstock & 1 & 0 & 0 & 0 & $\omega^2$ \\
Szabo & $\nu$-1 & 0 & 0 & 0 & $\omega^{\nu}$ \\
Stokes & -1 & 1 & 0 & 0 & $\omega^2$ \\
Caputo, Wismer & $\nu$-3 & 1 & 0 & 0 & $\omega^{\nu}$ \\
Chen and Holm & -1 & $\nu$/2 & 0 & 0 & $k^{\nu}$ \\
Treeby and Cox & -1 & $\nu$/2 & -2 & ($\nu$+1)/2 & $k^{\nu}$ \\
\hline
\end{tabular}
\end{table}
\subsubsection{Inhomogeneous medium}
Consider that a spherical region of radius $a$ exists within the computational domain and the speed of sound of this region differs from that of the surrounding medium. The inhomogeneous region will cause scattering of the impinging wave. The time dependent wave equation can be cast as \cite{Morse1, RKS3, RKS4}, 
\begin{eqnarray}
\nabla^2 \Psi(\textbf{r}, t) -\frac{1}{v_f^2} \frac{\partial^2}{\partial t^2} \Psi(\textbf{r}, t) = \frac{1}{v_f^2} \gamma_{\kappa}(\textbf{r}, t) \frac{\partial^2}{\partial t^2} \Psi(\textbf{r}, t)  + \nabla .[\gamma_{\rho}(\textbf{r}, t)\nabla \Psi(\textbf{r}, t)],
\label{acoustic_waveequn_time_dependent_inhomogeneous}
\end{eqnarray}
where 
\begin{eqnarray}
\gamma_{\kappa}(\textbf{r}, t)=
\frac{\kappa_s(\textbf{r}, t)-\kappa_f}{\kappa_f} ~~~~~ \mbox ~~~~~\gamma_{\rho}(\textbf{r}, t)=
\frac{\rho_s(\textbf{r}, t)-\rho_f}{\rho_s(\textbf{r}, t)},
\label{mismatch_parameters}
\end{eqnarray}
inside the scattering region; and $\gamma_{\kappa}(\textbf{r}, t)=\gamma_{\rho}(\textbf{r}, t)=0$ outside the scatterer; $\rho_f$ and $\kappa_f$ notations state the density and compressibility of the ambient medium whereas the same quantities for the scatterer are denoted by $\rho_s$ and $\kappa_s$, respectively; the subscript $s$ denotes the scattering region. For $\rho_s\approx \rho_f$, Eq. (\ref{acoustic_waveequn_time_dependent_inhomogeneous}) in the frequency domain can be approximated as,
\begin{eqnarray}
\nabla^2 \psi(\textbf{r}) + k_f^2 \psi(\textbf{r}) &=&  -(k_s^2-k_f^2)\psi(\textbf{r}), ~~~~~~~\mbox{inside the scatterer} \nonumber \\
\nabla^2 \psi(\textbf{r}) + k_f^2 \psi(\textbf{r}) &=&  0. ~~~~~~~\mbox{outside the scatterer} \nonumber \\
\label{acoustic_waveequn_time_independent_inhomogeneous}
\end{eqnarray}
As stated above, exact solutions of Eq. (\ref{acoustic_waveequn_time_independent_inhomogeneous}) can be deduced for regular shapes by imposing continuity of pressure and normal component of the particle velocity on the surface of the inhomogeneity \cite{Morse1, RKSPhDThs}
\subsubsection{Nonlinear medium}
The time-dependent wave equation for a nonlinear medium yields \cite{Hamilton},
\begin{eqnarray}
\nabla^2 \Psi(\mathbf{r}, t) - \frac{1}{c_0^2} \frac{\partial^2 \Psi(\mathbf{r}, t)}{\partial t^2} + \xi \frac{\partial^2 \Psi^2(\mathbf{r}, t)}{\partial t^2} = -S(\textbf{r}) \cos(\omega t)
\label{acoustic_waveequn_time_dependent_nonlinear}
\end{eqnarray}
where $\xi = \frac{1}{\rho_fv_f^4}(1+\frac{1}{2}\frac{B}{A})$ and $B/A$ is called the nonlinearity parameter, which characterizes the relative contribution of
finite-amplitude effects to the sound speed \cite{Hamilton}. The above equation closely resembles the non-absorbing Westervelt equation \cite{Hamilton}. The acoustic source is sinusoidally varying with time with an angular frequency $\omega$. Substituting $\Psi(\mathbf{r}, t) = \sum_{n=1}^{\infty} A_n\cos(n\omega t -\Phi_n)$ into Eq. (\ref{acoustic_waveequn_time_dependent_nonlinear}), one obtains the following coupled Helmholtz equations \cite{Kaltenbacher},
\begin{subequations}
\begin{eqnarray}
\nabla^2 \psi_1 + k_1^2 \psi_1 &=& - S(\textbf{r}) + \xi (1\omega)^2 \Big[~~~~~~~~~~~~~~~~~~~~~~\psi_1^*\psi_2+\psi_2^*\psi_3+\psi_3^*\psi_4+\psi_4^*\psi_5\Big],
\label{nonlinear_waveeqn_shi1}
\end{eqnarray}
\begin{eqnarray}
\nabla^2 \psi_2 + k_2^2 \psi_2 &=& ~~~~~~~~~~~  \xi (2\omega)^2 \Big[~~~~~~~~~~\frac{1}{2} \psi_1\psi_1+\psi_1^*\psi_3+\psi_2^*\psi_4+\psi_3^*\psi_5+\psi_4^*\psi_6\Big],
\label{nonlinear_waveeqn_shi2}
\end{eqnarray}
\begin{eqnarray}
\nabla^2 \psi_3 + k_3^2 \psi_3 &=& ~~~~~~~~~~~ \xi (3\omega)^2\Big[  ~~~~~~~~~~~~\psi_1\psi_2+\psi_1^*\psi_4+\psi_2^*\psi_5+\psi_3^*\psi_6+\psi_4^*\psi_7 \Big],
\label{nonlinear_waveeqn_shi3}
\end{eqnarray}
\begin{eqnarray}
\nabla^2 \psi_4 + k_4^2 \psi_4 &=& ~~~~~~~~~~~\xi (4\omega)^2 \Big[\frac{1}{2} \psi_2\psi_2+\psi_1\psi_3+\psi_1^*\psi_5+\psi_2^*\psi_6+\psi_3^*\psi_7+\psi_4^*\psi_8 \Big],
\label{nonlinear_waveeqn_shi4}
\end{eqnarray}
\begin{eqnarray}
\nabla^2 \psi_5 + k_5^2 \psi_5 &=& ~~~~~~~~~~~\xi (5\omega)^2 \Big[~~\psi_2\psi_3+\psi_1\psi_4+\psi_1^*\psi_6+\psi_2^*\psi_7+\psi_3^*\psi_8+\psi_4^*\psi_9\Big],
\label{nonlinear_waveeqn_shi5}
\end{eqnarray}
\label{acoustic_waveequn_time_independent_nonlinear}
\end{subequations}
where the superscript $*$ indicates the complex conjugate of a field. Eq. (\ref{nonlinear_waveeqn_shi1}) represents the wave propagation for the fundamental frequency. Eqs. (\ref{nonlinear_waveeqn_shi2}) to (\ref{nonlinear_waveeqn_shi5}) describe wave propagation for the 1st to 4th harmonics, respectively. Note that nonlinear propagation generates harmonics at multiples of the source frequency. The terms on the right hand side of Eq. (\ref{acoustic_waveequn_time_independent_nonlinear}) containing complex conjugates of the pressure fields are called the difference terms, rest are designated as the sum terms. A difference term has a frequency that is the difference between the frequencies of the components it arose from, a sum term has a frequency that is the sum of the frequencies of the components it arose from. The sum terms typically dominate, to the extent that difference terms are often neglected in practice. In other words, the accuracy of the field calculation can be improved by increasing the number of difference terms. Moreover, the field strength decreases as the frequency increases. 
\subsection{Derivation of the Born series}
The general form of the Helmholtz equation in presence of source terms for a lossless medium can be written as \cite{Osnabrugge},
\begin{eqnarray}
\nabla^2 \psi(\textbf{r}) + k_f^{2}  \psi(\textbf{r})   =  -S(\textbf{r})  - U(\textbf{r})\psi(\textbf{r}),
\label{acoustic_waveequn5}
\end{eqnarray}
where $k_f$ is the wavenumber, $S$ is a source term causing acoustic emission and and $U$, is the potential term appearing because of mismatach of acoustic properties of the region of interest with respect to the surrounding medium or for other factors such absorption and dispersion etc. The above equation can also be cast as \cite{Osnabrugge},
\begin{eqnarray}
\nabla^2 \psi(\textbf{r}) + (k_f^{2} + i \epsilon) \psi(\textbf{r})   =  -S(\textbf{r})  - V(\textbf{r})\psi(\textbf{r}),
\label{acoustic_waveequn6}
\end{eqnarray}
with $V(\textbf{r})=U(\textbf{r})-i\epsilon$, 
$\epsilon$ being an infinitesimally small real number. In Eq. (\ref{acoustic_waveequn6}), the propagating medium has been artificially made as an attenuating medium. However, the same attenuation factor has been included in the potential term which amplifies the field compensating the attenuation effect. The solution to Eq. (\ref{acoustic_waveequn6}) can be obtained as \cite{Morse}, 
\begin{eqnarray}
\psi(\textbf{r}) = \int g(\textbf{r}|\textbf{r}_0)[V (\textbf{r}_0)\psi(\textbf{r}_0) + S(\textbf{r}_0)] d^3\textbf{r}_0.
\label{soln_greens_fn}
\end{eqnarray}
where, $g(\textbf{r}|\textbf{r}_0)$ is the Green's function that satisfies the following equation,
\begin{eqnarray}
\nabla^2 g(\textbf{r}|\textbf{r}_0) + (k_f^{2} + i \epsilon) g(\textbf{r}| \textbf{r}_0) = -\delta(\textbf{r}-\textbf{r}_0),
\label{greens_fn}
\end{eqnarray}
here, $\delta(\textbf{r}-\textbf{r}_0)$ is the Dirac delta function; the functional forms of the Green's functions in 2D and 3D can be found in the literature \cite{Arfken}; $g(\textbf{r}|\textbf{r}_0)$ essentially provides the field at point $\textbf{r}$ due to a source point located at $\textbf{r}_0$ for a lossy medium. It is a lossy Green's function due to finite $\epsilon$ and as a result of that the total energy of the Green's function becomes finite and localized. Thus the Fourier transform of $g$ vis-{\' a}-vis $\psi$ can be calculated. In other words, the introduction of $\epsilon$ in Eq. (\ref{acoustic_waveequn6}) helps to evaluate the field in a finite computational domain. Eq. (\ref{soln_greens_fn}) involves convolution sums and in terms of matrices, it can be represented as,  
\begin{eqnarray}
\psi = GV\psi + GS,
\label{TBS_1}
\end{eqnarray}
where $G = \mathcal{F}^{-1}\tilde{g}(\textbf{p})\mathcal{F}$; $\mathcal{F}$ and $\mathcal{F}^{-1}$ are the forward and inverse Fourier transform operators, respectively; $\tilde{g}(\textbf{p})$ is the Fourier transform of $g(\textbf{r}|\textbf{r}_0)$ and is given by, %
$\tilde{g}(\textbf{p})=\frac{1}{(|\textbf{p}|^2-k_f^2-i\epsilon)}$
with $\textbf{p}$ is the Fourier transformed coordinates. 
Eq. (\ref{TBS_1}) can be recursively expanded yielding, 
\begin{eqnarray}
\psi = [1+ GV + GVGV + ... ]GS.
\label{TBS_2}
\end{eqnarray}
Eq. (\ref{TBS_2}) is known as the TBS and it converges if the spectral norm of $|GV| < 1$ \cite{Osnabrugge}. In other words, the infinite series converges for small objects with weak scattering potentials. The infinite sum in Eq. (\ref{TBS_2}) can be carried out in the following manner,
\begin{eqnarray}
\psi^{l+1} (\textbf{r}) = \mathcal{F}^{-1}[\tilde{g}(\mathbf{p}) \mathcal{F} [V(\mathbf{r}) \psi^{l}(\mathbf{r}) + S(\mathbf{r})]]. 
\label{TBS_algorithm}
\end{eqnarray}
Therefore, the field at the $(l+1)$th iteration is calculated using that of the $l$th step. The field at the $0$th iteration can be estimated as,
\begin{eqnarray}
\psi^{0} (\textbf{r}) = \mathcal{F}^{-1}[\tilde{g}(\textbf{p}) \mathcal{F} [S(\textbf{r})]].
\label{TBS_input}
\end{eqnarray}
It will act as the input to Eq. (\ref{TBS_algorithm}) and field would be computed successively as the iteration progresses. 
\subsection{Solutions to the wave equations using the
Born series method}
\subsubsection{Wave propagation through a homogeneous and  lossless medium} 
Eq. (\ref{acoustic_waveequn_time_independent}) after a minor modification becomes,
\begin{eqnarray}
\nabla^2 \psi(\textbf{r}) + (k_f^{2} + i \epsilon) \psi(\textbf{r})   =  -S(\textbf{r}) + i\epsilon \psi(\textbf{r}),
\label{TBS_equation_lossless}
\end{eqnarray}
here, $S(\textbf{r})$ term is introduced because the region of interest retains an acoustic source (e.g., transducer elements). The solution to Eq. (\ref{TBS_equation_lossless}) in terms of the TBS method appears as,
\begin{eqnarray}
\psi^{l+1} (\textbf{r}) = \mathcal{F}^{-1}[\tilde{g}(\mathbf{p})\mathcal{F}[-i\epsilon \psi^{l}(\mathbf{r}) + S(\mathbf{r})]],
\label{TBS_solution_lossless}
\end{eqnarray}
with $V(\textbf{r})=-i\epsilon$. All multiplications have to be performed element wise. Eq. (\ref{TBS_solution_lossless}) has been realized in this work to calculate the acoustic field generated by a collection of point transducer-elements. 
\subsubsection{Wave propagation through a homogeneous but absorbing and dispersive medium}
Similarly, Eq. (\ref{acoustic_waveequn_time_independent_absorbing_dispersion}) can be rewritten as,
\begin{eqnarray}
\nabla^2 \psi(\textbf{r}) + (k_f^2+i\epsilon) \psi(\textbf{r}) = -S(\textbf{r}) - \{-i \tau \omega (-\nabla^2)^{\nu/2} + \eta (-\nabla^2)^{(\nu+1)/2}-i\epsilon \}\psi(\textbf{r}),
\label{TBS_equation_absorbing-dispersive}
\end{eqnarray}
where the potential term is recognized as, $V(\textbf{r}) = -i \tau \omega (-\nabla^2)^{\nu/2} + \eta (-\nabla^2)^{(\nu+1)/2} - i \epsilon$. For this potential, the TBS procedure offers,
\begin{eqnarray}
\psi^{l+1} (\textbf{r}) = \mathcal{F}^{-1}[\tilde{g}(\mathbf{p})[\{-i \tau \omega k^{\nu} + \eta  k^{\nu+1}-i\epsilon\}\mathcal{F}[\psi^{l}(\mathbf{r})]+\mathcal{F}[S(\mathbf{r})]]]. 
\label{TBS_solution_absorbing-dispersive}
\end{eqnarray}
In this derivation, we have applied $\mathcal{F}[(-\nabla^2)^{{\nu}/2}\psi^{l} (\mathbf{r})]=k^{\nu}\mathcal{F}[\psi^{l}(\mathbf{r})]$. The above expression has been computed to determine the steady state acoustic field generated by a linear array of point emitters when the medium is absorbing and dispersive. 
\subsubsection{Wave propagation through an inhomogeneous medium} 
Eq. (\ref{acoustic_waveequn_time_independent_inhomogeneous}), after adding $i\epsilon \psi$ both sides, transforms into, 
\begin{eqnarray}
\nabla^2 \psi(\textbf{r}) + (k_f^2+i\epsilon) \psi(\textbf{r}) = -\mathcal{S}(\textbf{r}) -(k_s^2-k_f^2-i\epsilon)\psi(\textbf{r})
\label{TBS_equation_inhomogeneous}
\end{eqnarray}
The potential term can readily be identified as,
$V(\textbf{r})=k_s^2-k_f^{2} - i \epsilon$ within the scatterer and 
$V(\textbf{r})=-i\epsilon$ in the ambient medium. Accordingly, one obtains,
\begin{subequations}
\begin{eqnarray}
\psi^{l+1} (\textbf{r}) = \mathcal{F}^{-1}[\tilde{g}(\mathbf{p})\mathcal{F}[(k_s^2-k_f^2-i\epsilon) \psi^{l}(\mathbf{r}) + S(\mathbf{r})]],~~~~\mbox{inside the scatterer}
\label{TBS_solution_inhomogeneous_inside}
\end{eqnarray}
\begin{eqnarray}
\psi^{l+1} (\textbf{r}) = \mathcal{F}^{-1}[\tilde{g}(\mathbf{p})\mathcal{F}[-i\epsilon \psi^{l}(\mathbf{r}) + S(\mathbf{r})]],~~~~\mbox{outside the scatterer}
\label{TBS_solution_inhomogeneous_outside}
\end{eqnarray}
\label{TBS_solution_inhomogeneous}
\end{subequations}
Eq. (\ref{TBS_solution_inhomogeneous}) is solved herein to evaluate the pressure field provided by an array of tiny transducer-segments in presence of a scatterer. 
\subsubsection{Wave propagation through a nonlinear medium}
Following the same step, Eq. (\ref{acoustic_waveequn_time_independent_nonlinear}) can be presented as,
\begin{subequations}
\begin{eqnarray}
\nabla^2 \psi_1 + (k_1^2+i\epsilon_1) \psi_1 &=& -S(\mathbf{r})~~~~~~~~~~~~~~~~~~~~~~+i\epsilon_1 \psi_1,
\label{TBS_equation_nonlinear_sum_terms_psi1}
\end{eqnarray}
\begin{eqnarray}
\nabla^2 \psi_2 + (k_2^2+i\epsilon_2) \psi_2 &=& \xi (2\omega)^2 [~~~~~~~~~~\frac{1}{2} \psi_1\psi_1]+i\epsilon_2\psi_2,
\label{TBS_equation_nonlinear_sum_terms_psi2}
\end{eqnarray}
\begin{eqnarray}
\nabla^2 \psi_3 + (k_3^2+i\epsilon_3) \psi_3 &=& \xi (3\omega)^2 [~~~~~~~~~~~~\psi_1\psi_2]+i\epsilon_3\psi_3,
\label{TBS_equation_nonlinear_sum_terms_psi3}
\end{eqnarray}
\begin{eqnarray}
\nabla^2 \psi_4 +  (k_4^2+i\epsilon_4) \psi_4 &=& \xi (4\omega)^2 [\frac{1}{2} \psi_2\psi_2+\psi_1\psi_3]+i\epsilon_4\psi_4,
\label{TBS_equation_nonlinear_sum_terms_psi4}
\end{eqnarray}
\begin{eqnarray}
\nabla^2 \psi_5 + (k_5^2+i\epsilon_5) \psi_5 &=& \xi (5\omega)^2 [ ~~\psi_2\psi_3+\psi_1\psi_4]+i\epsilon_5\psi_5.
\label{TBS_equation_nonlinear_sum_terms_psi5}
\end{eqnarray}
\label{TBS_equation_nonlinear_sum_terms}
\end{subequations}
For simplicity, only the sum terms are retained in Eqs. (\ref{TBS_equation_nonlinear_sum_terms_psi1}) to (\ref{TBS_equation_nonlinear_sum_terms_psi5}). The difference terms may have less contributions and thus, have been ignored. The source and potential terms, as per Eq. (\ref{acoustic_waveequn6}), can be distinguished without any difficulty for each equation. The corresponding solutions under the TBS framework take the forms,
\begin{subequations}
\begin{eqnarray}
\psi_1^{l+1} (\textbf{r}) &=& \mathcal{F}^{-1}[\tilde{g_1}(\mathbf{p})\mathcal{F}[-i\epsilon_1 \psi_1^{l} + S]]. 
\label{TBS_solution_shi_1}
\end{eqnarray}
\begin{eqnarray}
\psi_2^{l+1} (\textbf{r}) &=& \mathcal{F}^{-1}[\tilde{g_2}(\mathbf{p})\mathcal{F}[-i\epsilon_2\psi_2^{l} -\xi (2\omega)^2 \frac{1}{2} \psi_1\psi_1]]. 
\label{TBS_solution_shi_2}
\end{eqnarray}
\begin{eqnarray}
\psi_3^{l+1} (\textbf{r}) &=& \mathcal{F}^{-1}[\tilde{g_3}(\mathbf{p})\mathcal{F}[-i\epsilon_3\psi_3^l -\xi (3\omega)^2  \psi_1\psi_2]]. 
\label{TBS_solution_shi_3}
\end{eqnarray}
\begin{eqnarray}
\psi_4^{l+1} (\textbf{r}) &=& \mathcal{F}^{-1}[\tilde{g_4}(\mathbf{p})\mathcal{F}[ -i\epsilon_4\psi_4^l -\xi (4\omega)^2 [\frac{1}{2} \psi_2\psi_2+\psi_1\psi_3]]]. 
\label{TBS_solution_shi_4}
\end{eqnarray}
\begin{eqnarray}
\psi_5^{l+1} (\textbf{r}) &=& \mathcal{F}^{-1}[\tilde{g_5}(\mathbf{p})\mathcal{F}[-i\epsilon_5\psi_5^l -\xi (5\omega)^2 [ \psi_2\psi_3+\psi_1\psi_4]]]. 
\label{TBS_solution_shi_5}
\end{eqnarray}
\label{TBS_solution_nonlinear_shi_n}
\end{subequations}
where $\tilde{g}_n(\textbf{p})=\frac{1}{(|\textbf{p}|^2-n^2k_{f}^2-i\epsilon_n)}$. The series of equations furnished above have been numerically evaluated to estimate the pressure fields produced by the fundamental and higher harmonics for a nonlinear medium.
\section{Numerical procedures}
\subsection{Implementation of the traditional Born series method} 
\begin{figure*}[!h]
\centerline{\includegraphics[width=1.0\textwidth]{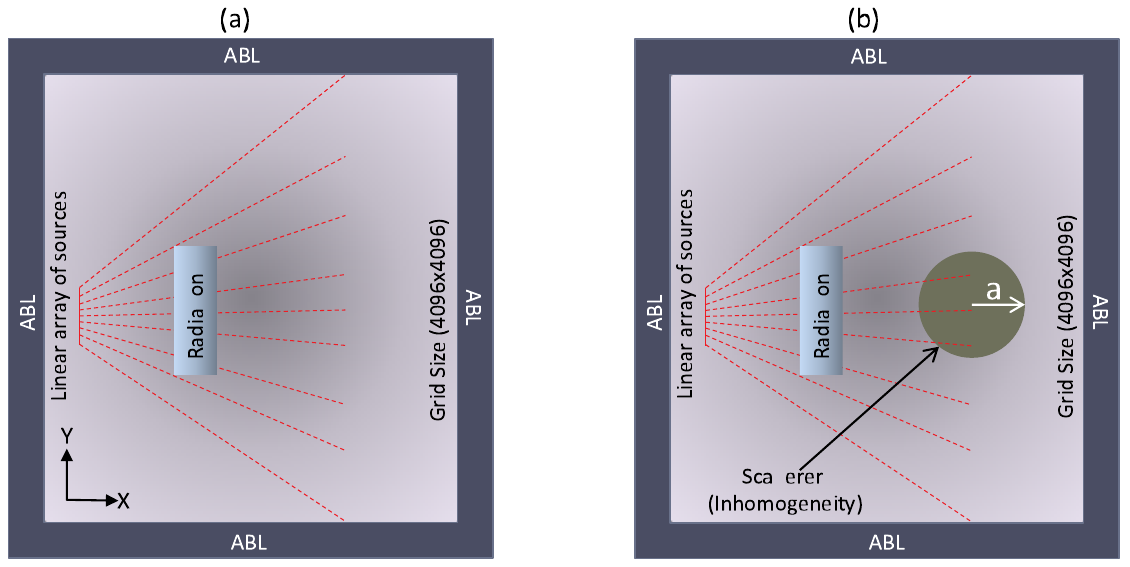}}
\caption{Schematic diagram of the computational domain. (a) The propagating medium is homogeneous but either lossless or lossy (absorbing/absorbing-dispersive) or nonlinear. (b) The propagating medium encloses a scattering region. ABL refers to absorbing layer.}
\label{fig_1}
\end{figure*}

An algorithm realizing the TBS procedure for linear media is presented in Algorithm \ref{alg:TBSalgorithm_linear_nonlinear}. The computational domain consisted of $4096\times4096$ pixels with sides $dx=dy=0.02\times10^{-3}$ m. Therefore, different matrix operations like initialization, addition, multiplication were performed on $4096\times4096$ matrices. The speed of sound and density of the medium were fixed to, $v_f=1500$ m/s and $\rho_f=1000$ kg/m$^3$. A representative diagram is shown in Fig. \ref{fig_1}. A linear array of acoustic sources was placed at a distance $x_{el}=675dx$ from the left edge as shown in Fig. \ref{fig_1} with a strength of $|S|=1000000$. The source points were symmetrically 
positioned with respect to the row number $y_{el}=2049dx$. A total number of $N_{el}=501$ grid points (transducer elements) were acted as the acoustic sources replicating a line source of width $500dx$. The initial field was estimated by implementing Eq. (\ref{TBS_input}). After that, for the homogeneous and lossless medium, Eq. (\ref{TBS_solution_lossless}) was calculated iteratively to obtain the spatial distribution of pressure field for $f=1007080$ Hz using $\epsilon=0.8k_f^2$ \cite{Osnabrugge, Anuj_JMO, RKS1, RKS2, UM_Photoacoustics}. The chosen frequency $f=1007080$ Hz is the 55th multiple of the fundamental frequency supported by this grid system ($v_f/(4096dx)$). Note that after each iteration the pressure field $\psi^{l+1}$ was multiplied by a Sigmoid function \cite{UM_Photoacoustics},
\begin{eqnarray}
f_{\mbox{ABL}}(x)=\frac{1}{1+\exp({-\zeta (x-0.5\mu)})}, \text{within the ABL},
\label{ABL_profile}
\end{eqnarray}
%
% %
% %
where $x>0$; $\zeta$ was chosen to be, $\zeta=21.48$ Np/cm for all frequencies. This function essentially built the absorbing layer (ABL), which was placed at the boundaries of the computational domain as could be seen from Fig. \ref{fig_1}. The width of the ABL was $\mu=450dx$. It was designed with a nonlinear absorption property. This layer ensured that an outgoing wave was absorbed effectively and exiting a boundary did not reemerge from the opposite side, preventing wave wrapping associated with the Fourier transformation operation \cite{Treeby1}. The field was supposed to converge if the error was $<10^{-5}$. The error was obtained as,
\begin{eqnarray}
\mbox {Total error} = \frac{\sum_{m=1}^{4096}|\psi_{l+1}(2049, m)-\psi_{l}(2049, m)|}{\sum_{m=1}^{4096}|\psi_{l}(2049, m)|}.
\label{Error}
\end{eqnarray}
A maximum number of 2000 iterations was set in all simulations. An algorithm applying the TBS scheme in the context of photoacoustics can be found in \cite{Anuj_JMO}. MATLAB implementations of the Born series methods are available in \cite{github}.

Similarly, Eq. (\ref{TBS_solution_absorbing-dispersive}) was implemented to yield the field distribution for the same transducer in an absorbing and dispersive medium having $\alpha_0=5.5 \times 10^{-10}$ Np(rad/s)$^{\nu}$ m$^{-1}$ and $\nu=1.5$ \cite{Treeby1}. Accordingly, $\tau$ and $\eta$ could be computed to be $\tau=-2\alpha_0v_f^{\nu-1}$ and $\eta=2\alpha_0v_f^{\nu} \tan(\pi \nu/2)$. The initial field was determined as that of the previous case. For the inhomogeneous case, Eq. (\ref{TBS_solution_inhomogeneous}) was numerically realized by repeating the same procedure. The speed of sound inside the sactterer was varied from $v_s$ = 1350 to 1800 m/s with a step of 150 m/s. The radius of the inhomogeneity was taken as $a = 50dx = 1$ mm with size parameter $k_fa=4.22$. It is positioned at $x=40.2$ mm away from the center of the acoustic line source. 

Finally, pressure fields for a nonlinear medium for the same transducer were assessed for the fundamental frequency as well as up to the 5th harmonic. The wave equations are given in Eqs. (\ref{TBS_equation_nonlinear_sum_terms}) and the corresponding TBS solutions are provided in Eqs. (\ref{TBS_solution_nonlinear_shi_n}). The TBS algorithm implemented for this purpose is detailed in Algorithm \ref{alg:TBSalgorithm_linear_nonlinear}. At first, the pressure field for the fundamental frequency was evaluated. After that fields were calculated for higher harmonics one by one. The numerical values of $\epsilon_n=0.8n^2k_f^2$ and $\tilde{g}_n(\textbf{p})=\frac{1}{(|\textbf{p}|^2-n^2k_{f}^2-i\epsilon_n)}$ were appropriately calculated for each harmonic \cite{Osnabrugge, Anuj_JMO, RKS1, RKS2, UM_Photoacoustics}; $B/A$ was assigned to be 5 \cite{Beyer}. Note that the initial field for a harmonic was obtained from suitable combinations of estimated fields at lower harmonics. 

The pressure field distributions for various media were calculated by running the TBS codes written in the CUDA C programming language. All simulations were performed on a workstation using the GPU architecture. The specifications of the machine are included in Table \ref{tab:HPC_specs}. 
\begin{table}[!t]
\centering
\caption{Description of the computational resources used for numerical calculations.}
\label{tab:HPC_specs}
\begin{tabular}{|p{0.4\linewidth}|p{0.5\linewidth}|}
\hline
\rowcolor{gray!30}\textbf{Feature} & \textbf{Description} \\
\hline
CPU Model & AMD Ryzen Threadripper PRO 5965WX 24-Cores @4569.0000 MHz \\
Number of CPU Cores & 24 \\
Thread(s) per core & 2 \\
Core(s) per socket & 24 \\
Architecture & x86\_64, 64-bit \\
Operating System (OS) & Ubuntu 24.04.2 LTS \\
GPU Model & NVIDIA GeForce RTX 4060 @2.475 GHz \\
Number of CUDA cores & 3072 \\
GPU Driver Version & 570.124.06 \\
Number of GPUs & 1 \\
GPU RAM & 8 GB  \\
CPU RAM & 256 GB\\
CUDA Toolkit Version & 12.8 \\
MATLAB Version & 2023b \\
\hline
\end{tabular}
\end{table}
%

%%%%%%%%%%%%%%% TBS Algorithm for linear and nonlinear media %%%%%%%%%%%%%%%%%%%%%%
%
\begin{algorithm}[]  
\caption{Acoustic field calculation using the TBS algorithm for linear media.}
\label{alg:TBSalgorithm_linear_nonlinear}
\DontPrintSemicolon
\SetAlgoLined
\SetKwInOut{Input}{Input}
\SetKwInOut{Output}{Output}

\Input{
Initialize system parameters 
Prepare computational domain: $N_{cn}$, $K_{cn}$, $dx$, $ABL$ 
Convergence limit: $ThError$ 
Transducer properties: $f$, $N_{el}$, $x_{el}$, $y_{el}$ 
Medium properties: $v_f$, $\alpha$, $\tau$, $\eta$, $\nu$, $B/A$ 
Scatterer properties: $v_s$, $a$
}

\Output{PA field $\psi_{fn}$}

\For{$i \gets 1$ \KwTo $5$}{ \tcp*{Restrict $i=1$ for linear media}
  $f \gets i\times 55 \times v_f / (N_{cn} \times dx)$\; 
  $\omega \gets 2\pi f$\;
  $k_s \gets \omega / v_s$\; 
  $k_f \gets \omega / v_f$\; 
  $\epsilon \gets 0.8 k_f^2$\;
  
  \For{$j \gets 0$ \KwTo $N_{cn} - 1$}{
    \For{$m \gets 0$ \KwTo $N_{cn} - 1$}{
      $index \gets j \times N_{cn} + m$\;       
      $k_x \gets 2\pi (m - K_{cn}) / (N_{cn} dx)$\; 
      $k_y \gets 2\pi (j - K_{cn}) / (N_{cn} dx)$\;
      $kmod[index] \gets \sqrt{k_x^2 + k_y^2}$\; 
      $G[index] \gets \left(\frac{k_x^2 + k_y^2 - k_f^2}{(k_x^2 + k_y^2 - k_f^2)^2 + \epsilon^2}, \frac{\epsilon}{(k_x^2 + k_y^2 - k_f^2)^2 + \epsilon^2}\right)$\;
      
      \eIf{$(i==1)$}{
        \If{$(j==x_{el} \,\&\&\, m==y_{el})$}{
          $S[index] \gets (0.0, -1\times10^6)$\;
        }{
          $S[index] \gets (0.0, 0.0)$\;
        }
      }{
        $S[index] \gets \xi(\omega)^2\times \text{(sum terms)}$ \tcp*{From Eq. \ref{TBS_solution_nonlinear_shi_n}}
      }
      
      \tcp{Case I: Homogeneous medium}
      $V[index] \gets (0.0, -\epsilon)$\;
      
      \tcp{Case II: Absorbing/dispersive medium}
      $VV[index] \gets (\eta kmod^{\nu+1}, -\tau \omega kmod^{\nu}-\epsilon)$\;
      
      \tcp{Case III: Heterogeneous medium}
      $dist \gets \sqrt{(m - K_{cn})^2 + (j - K_{cn})^2} dx$\;
      \If{$dist \leq a$}{
        $V[index] \gets (k_s^2 - k_f^2, -\epsilon)$\;
      }\Else{
        $V[index] \gets (0.0, -\epsilon)$\;
      }
      
      \tcp{Case IV: Nonlinear medium}
      $V[index] \gets (0.0, -\epsilon)$\;    
    }
  }
  
  $G \gets \text{fftShift}(G)$\; 
  $kmod \gets \text{fftShift}(kmod)$\; 
  $VV \gets \text{fftShift}(VV)$\; 
  $\psi_{in} \gets \text{ifft}(G \times \text{fft}(S))$\;
  
  \For{$iter \gets 1$ \KwTo $2000$}{
    \tcp{Case I, III, IV:}
    $\psi_{fn} \gets \text{ifft}(G \times \text{fft}(V \times \psi_{in} + S)) \times ABL$\;
    \tcp{Case II:}
    $\psi_{fn} \gets \text{ifft}(G \times (VV \times \text{fft}(\psi_{in}) + \text{fft}(S))) \times ABL$\;
    
    $error \gets \frac{\|\psi_{fn}[N_{cn}/2, :] - \psi_{in}[N_{cn}/2, :]\|}{\|\psi_{in}[N_{cn}/2, :]\|}$\;
    \If{$error < ThError$}{
      $saturationTBS \gets iter$\;
      \textbf{break}\;
    }{
      $\psi_{in} \gets \psi_{fn}$\;
    }
  }
}
\end{algorithm}
\subsection{Computations using the k-Space pseudospectral technique}
The numerically evaluated pressure fields facilitated by the TBS approach were compared with that of a standard, MATLAB-based and freely available software (k-Wave toolbox). It is a widely accepted numerical tool for solving wave propagation problems, particularly in heterogeneous and nonlinear media. It provides accurate time-domain solution to the full wave equation. Comparison with k-Wave validates the accuracy and effectiveness of the TBS. However, other numerical methods could also be utilized for the same. The theoretical background of the k-Wave module and its implementation details can be found in the literature \cite{kwavetoolbox}, \cite{Treeby1}. For the sake of completeness, this procedure is briefly discussed here. In this approach, instead of second order partial differential equation, coupled first order partial differential equations are numerically solved. The most general form of the coupled partial differential equations incorporating all medium dependent scenarios are \cite{kwavetoolbox},
\begin{subequations}
\begin{eqnarray}
\frac{\partial \textbf{u}}{\partial t}=-\frac{1}{\rho_f}\nabla \psi + S_F,
\label{momemtum_c0nservation}
\end{eqnarray}
\begin{eqnarray}
\frac{\partial \rho}{\partial t}=-(2\rho+\rho_f)\nabla \cdot \textbf{u}-\textbf{u}\cdot\nabla \rho_f + S_M,
\label{mass_c0nservation}
\end{eqnarray}
\begin{eqnarray}
\psi = v_f^2\left ( \rho + \textbf{d} \cdot\nabla + \frac{B}{2A}\rho_f \frac{\rho^2}{\rho_f} - L_{frac}  \right )
\label{pressue_density_relation}
\end{eqnarray}
\label{kWave_equations}
\end{subequations}
here, the fractional Laplacian is defiend as, $L_{frac}=\tau \frac{\partial}{\partial t} (-\nabla^2)^{\frac{\nu}{2}-1} + \eta (-\nabla^2)^{\frac{\nu+1}{2}-1}$; $\textbf{u}$ is the acoustic particle velocity, $\rho$ is the density. Further, $S_F$ is a force source term and represents the input of body forces per unit mass and $S_M$ is a mass source term and represents the time rate of the input of mass per unit volume. Note that in the preceding wave equations, the mass source term has been considered (ignoring the force source term). Eqs. (\ref{kWave_equations}) demonstrate the momemtum conservation, mass conservation and pressure density relation, respectively. The above differential equations are efficiently calculated in the kWave module, using a pseudospectral model that computes spatial gradients using Fourier collocation and the time-stepping uses a dispersion-corrected finite difference approach. Additionally, staggering is used in space and time for reasons of stability and to improve accuracy for heterogeneous simulations.

The size of the computational domain and the grid spacing were the same as those of the previous simulations ($4096 \times 4096$, $dx=dy=0.02\times10^{-3}$ m). The thickness of the perfectly matched layer was chosen to be $30dx$. A linear array of acoustic sources were placed at the 40th grid-location from the outermost left boundary [see Fig. \ref{fig_1}]. The medium properties were set case by case as described above. The Courant–Friedrichs–Lewy number was fixed to 0.3 for the k-Wave simulations to ensure numerical stability. The time step was automatically fixed by the k-Wave toolbox. Then the GPU version of the k-Wave function, namely, kspaceFirstOrder2D was executed in the same workstation \cite{kwavetoolbox}. The steady state pressure data for three periods for individual grid locations were acquired. The time series data for each grid point were then Fourier transformed and subsequently, numerical values of amplitude and phase at $f=1007080$ Hz were stored (from the Fourier transformed data). This procedure was reiterated for all grid crossings. The next step was to generate a 2D map of amplitude and phase for comparison with the TBS counterparts. 
The above workflow was exactly followed for the nonlinear case as well but in this case, amplitude and phase at multiple frequencies, i.e., $f=1007080$, $2014160$, $3021240$, $4028320$ and $5035400$ Hz, were saved from the spectral data. 
\begin{figure*}[!t]
\centerline{\includegraphics[width=1.0\textwidth]{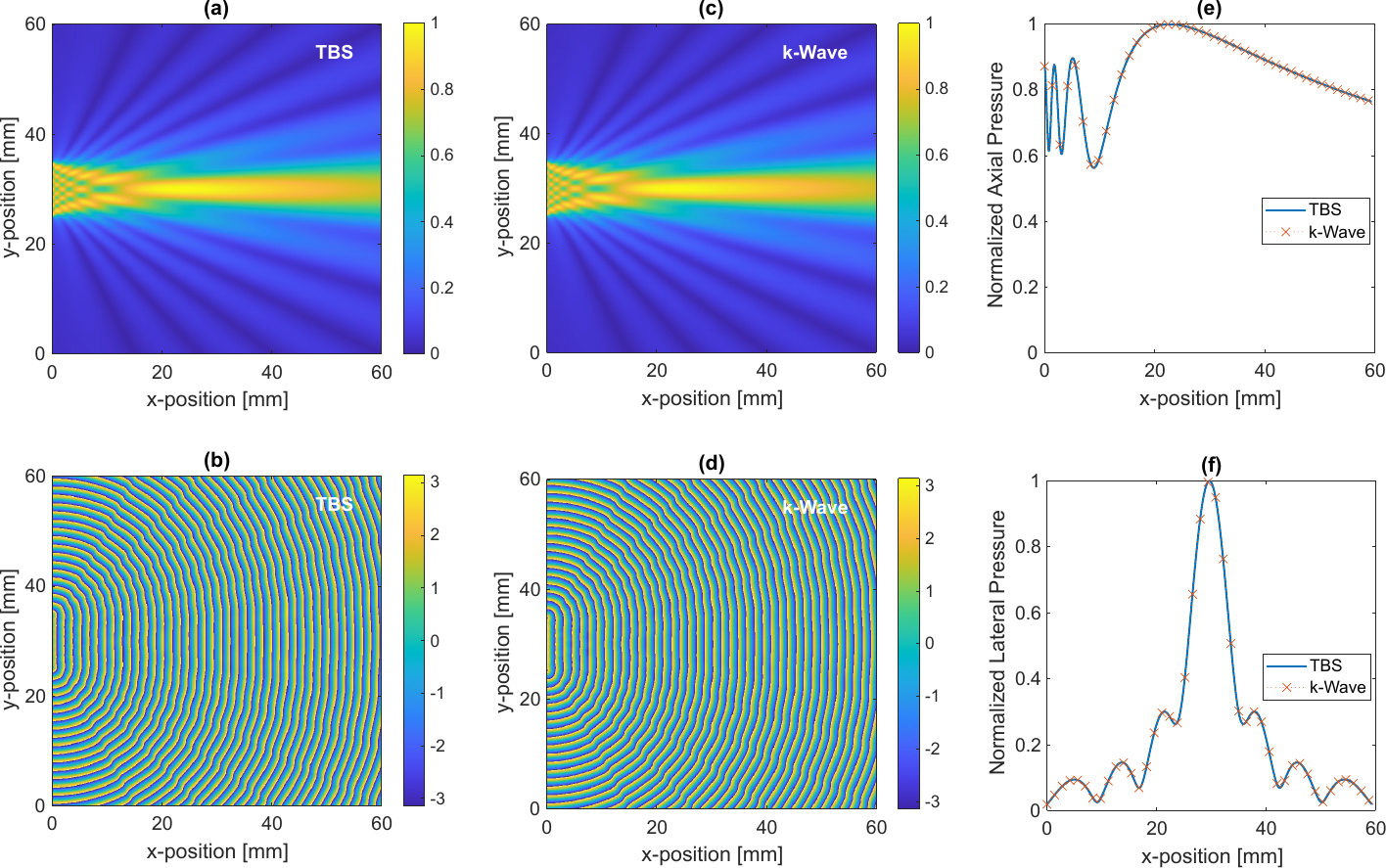}}
\caption{(a) and (b) Steady state normalized field pattern and spatial distribution of phase for a linear array source with 501 point elements calculated using the TBS method for homogeneous and lossless medium. (c) and (d) Same as (a) and (b) but for the k-Wave technique, respectively. (e) Steady-state acoustic normalized pressure amplitude along the axis of the transducer. (f) Same as (e) but for the lateral direction, at a distance $x\approx 41$ mm from the transducer.}
\label{Homogeneous}
\end{figure*}
\begin{figure*}[!h]
\centerline{\includegraphics[width=1.0\textwidth]{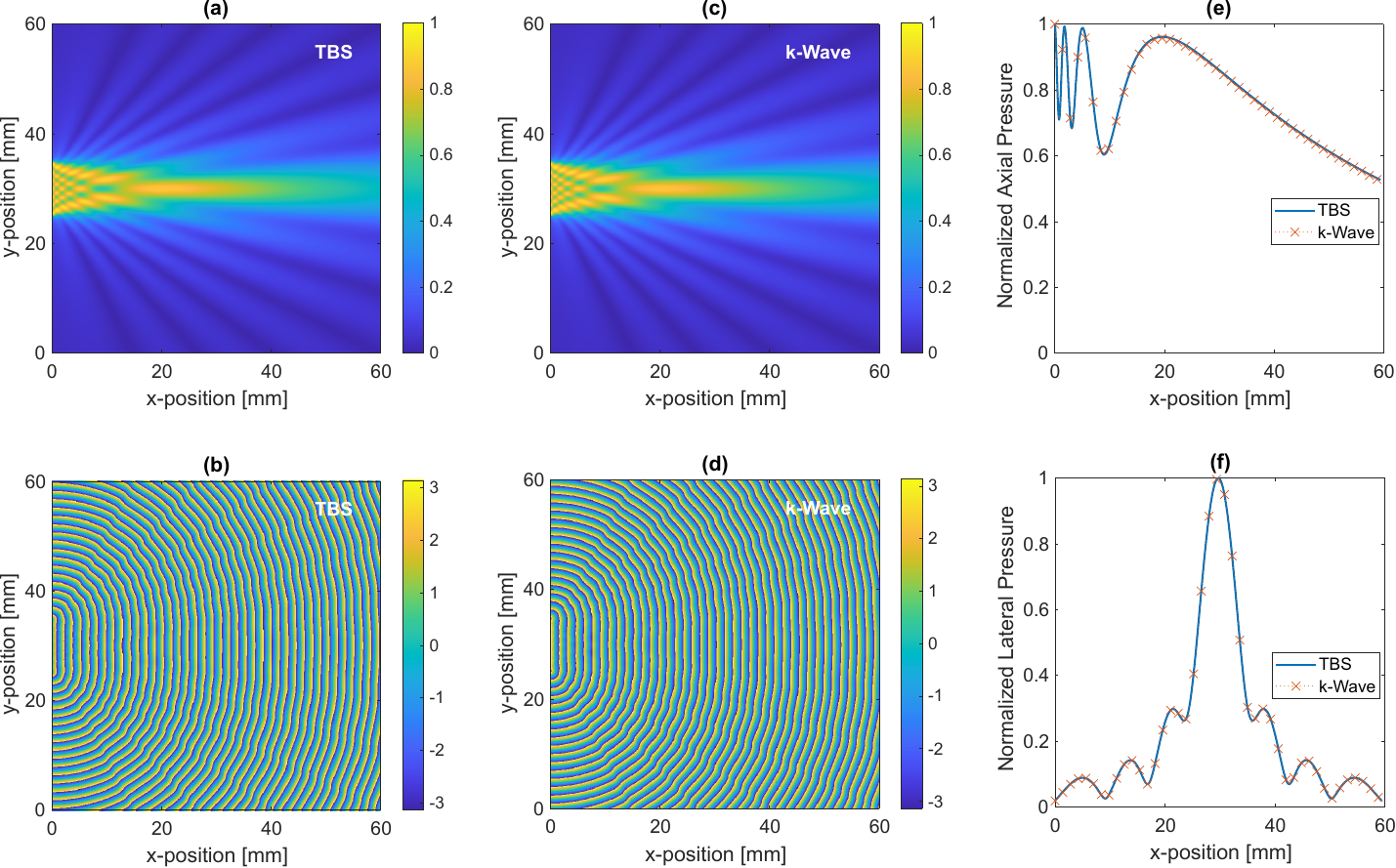}}
\caption{Same as Fig. \ref{Homogeneous} but for the absorbing and dispersive medium.}
\label{Absorbing_Dispersive}
\end{figure*}
\begin{figure*}[!t]
\centerline{\includegraphics[width=1.0\textwidth]{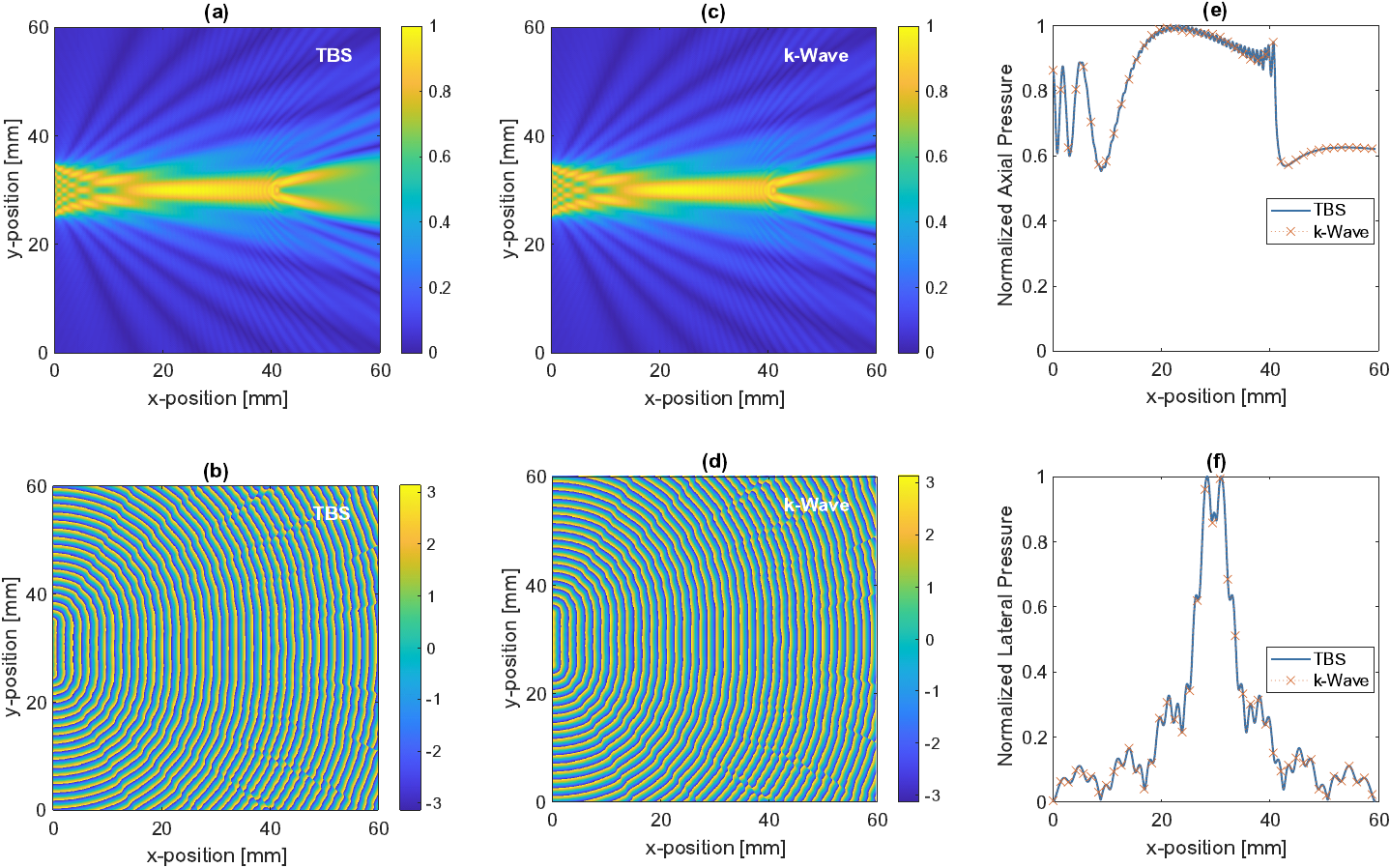}}
\caption{ (a) and (b) Steady state spatial distribution of normalized pressure amplitude and phase for the same linear array of sources and calculated using the TBS method for an inhomogeneous medium ($v_f=1500$ m/s, $v_s=1650$ m/s). The scatterer (of radius $a=1$ mm) and the transducer is separated by a distance of $x=40.2$ mm.  (c) and (d) Simulated results for the k-Wave procedure as of (a) and (b), respectively. (e) and (f) Plots of on and off axis profiles for the normalized pressure amplitude (for off axis distance from the line source is $x\approx 41$ mm).}
\label{Inhomogeneous_1650}
\end{figure*}
\begin{figure*}[!h]
\centerline{\includegraphics[width=1.0\textwidth]{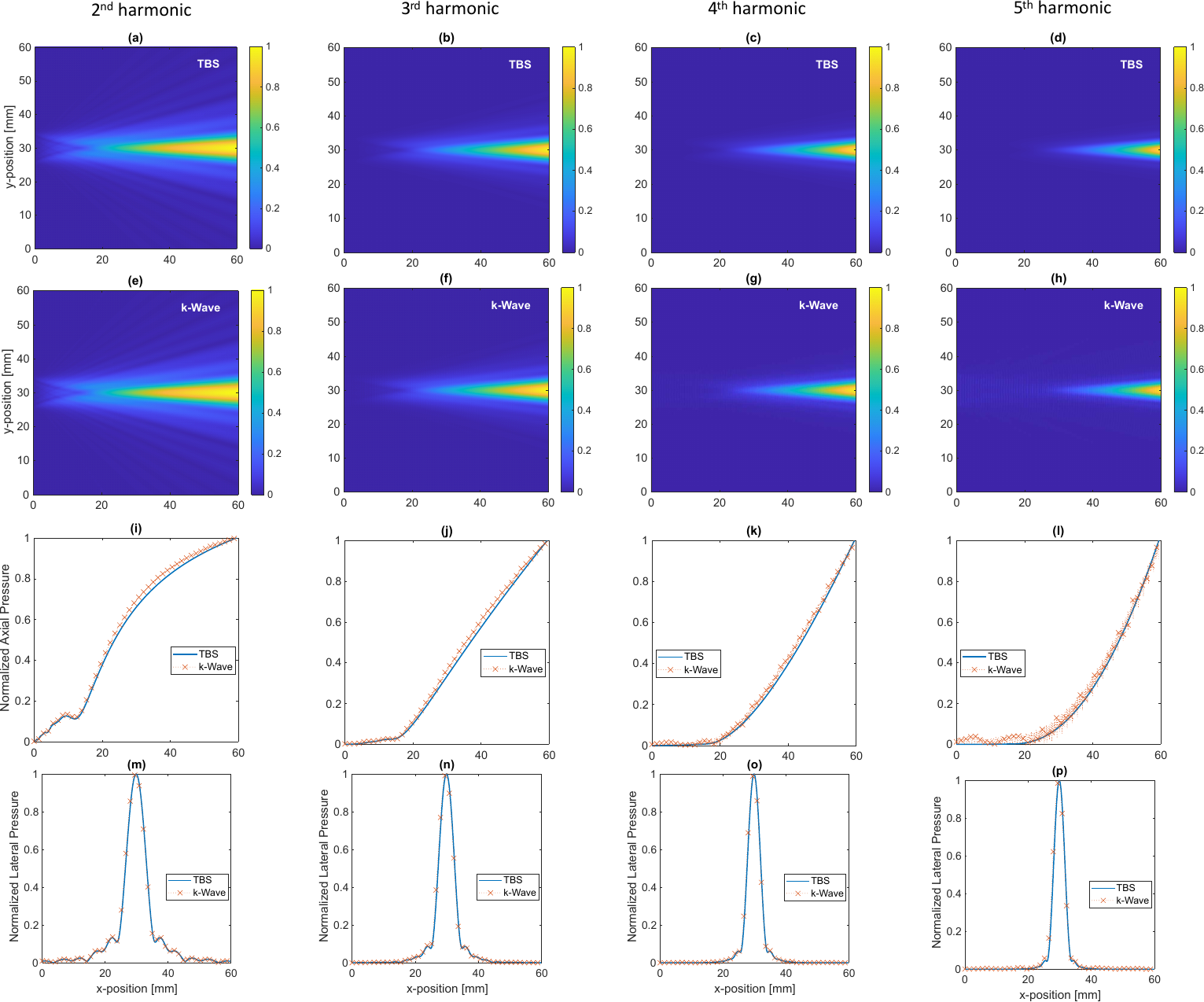}}
\caption{(a)-(d) Steady state normalized pressure field estimated by the TBS algorithm for 1st, 2nd, 3rd and 4th harmonics. (e)-(h) Analogous plots but simulated via the k-Wave package. (i)-(l) On axis variation of the steady state normalized pressure amplitude for the TBS and k-Wave schemes. (m)-(p) Demonstration of the off axis normalized pressure amplitude by the same methods at $x\approx 58$ mm from the transducer.} 
\label{Nonlinear}
\end{figure*}
%
%

%%%%%%%%%%%%%%
\section{Numerical results}
Figure \ref{Homogeneous} presents the simulated fields using the TBS and k-WAve methods for a linear array of point sources. The point sources are sinusoidally oscillating with a frequency of $f=1007080$ Hz. The propagating medium is  homogeneous and lossless. Perfect agreement can be seen between the two approaches. As expected, field near to the line source exhibits fluctuations owing to the diffraction effect (near field). In the far field, the main lobe dominates and it varies smoothly. In other words, most of the energy of the beam is packaged within a small angular region. The existence of the side lobes can also be noticed from Figs. \ref{Homogeneous} (a) and (c). The phase part changes between $-\pi$ to $\pi$. The on and off axes profiles look identical for the two approaches. 

Similar figures for the absorbing and dispersive medium are shown in Fig. \ref{Absorbing_Dispersive}. As in the previous case, both the procedures demonstrate exact match. It may be noted that pressure amplitude in the far field rapidly falls off for the main lobe due to the absorption properties of the medium. Compared to the lossless case, ultrasonic beam penetrates less in the lossy medium. The same plots for the absorbing medium are included in Fig. \ref{Absorbing}.

The spatial distributions of normalized pressure amplitude, phase and line plots are shown in Fig. \ref{Inhomogeneous_1650} when the medium contains a scattering region. The scatterer has a diameter of $a=1$ mm. It is placed on the axis of the line source and $x=40.2$ mm away from the acoustic source, where the seems to modify. The speed of sound mismatch is about 10\% with respect to that of the ambient medium ($v_s=1650$ m/s, $v_f=1500$ m/s). TBS and k-Wave results appear identical. The field pattern up to the source location is analogous to that of the homogeneous and lossless medium [compare Figs. \ref{Inhomogeneous_1650}(a) and \ref{Homogeneous}(a)]. After the inhomogeneity, the most of the beam energy splits into two angular directions (symmetrical with respect to the axis). Additional results for $v_s=1350$ and $1800$ m/s are shown in Figs. \ref{Inhomogeneous1350} and \ref{Inhomogeneous1800}, respectively.

The simulated field patterns for the nonlinear medium are displayed in Fig. \ref{Nonlinear} in details.  The field distributions created by the TBS methods are shown in the top row sequentially from 2nd to 5th harmonics, i.e.,  $f=2014160$, $3021240$, $4028320$ and $5035400$ Hz [Figs. \ref{Nonlinear}(a)-(d)]. The same results produced by the k-Wave are portrayed in the second row. The TBS result duplicates k-Wave estimations. It is evident from this figure that the region with maximum pressure amplitude shifts towards right (i.e., away from the source) as the frequency grows. The axial and lateral line profiles are drawn and depicted in the third and fourth rows, respectively. Up to third harmonics profiles are almost overlapping. However, for the fourth and fifth harmonics, k-Wave results deviate from that of the TBS technique. Reflections from the boundary occur for these harmonics [see Figs. \ref{Nonlinear}(g) and (h)]. 
\begin{table}[h]
    \centering
    \caption{Number of iterations and execution time required to obtain converging solutions.}
    \label{tab:execution_time}
    \renewcommand{\arraystretch}{1.6} 
    \begin{tabular}{p{3cm} c c c c p{3cm}}
        \toprule
        \textbf{Medium} & \textbf{TBS GPU} & \textbf{TBS GPU}  & \textbf{TBS CPU} & \textbf{TBS CPU} & \textbf{k-Wave} \\
             & \textbf{Iterations} & \textbf{Exe. Time (s)} & \textbf{Iterations} & \textbf{Exe. Time (s)} & \textbf{Exe. Time (s)} \\
        \midrule        
        Homogeneous                & 135  & 5 & 133 &  21 & 502.14 \\ 
        Absorbing                    & 133 & 8 & 134 &  31 &  680.81 \\ 
        Absorbing-dispersive         &  132 & 11 & 131 & 32 &  516.36 \\ 
        Inhomogeneous, $v_s = 1350$ m/s & 134 & 7 & 162 & 27&  551.94 \\ 
       Inhomogeneous, $v_s = 1650$ m/s & 134 & 7 &  155 & 25 & 558.25 \\ 
       Inhomogeneous, $v_s = 1800$ m/s & 134 & 7 & 164 &  27 & 607.23 \\ 
       Nonlinear, 1st harmonic/fundamental  &  135&  5 & 133 &  21 &     499.65\\ 
       Nonlinear, 2nd harmonic  &  228&  9& 225 & 34 &   N/A\\
       Nonlinear, 3rd harmonic  &  306&  12& 301 & 45 &   N/A\\
       Nonlinear, 4th harmonic  &  368&  15&  362 & 55 &  N/A\\
       Nonlinear, 5th harmonic  &  416&  16&  409 & 61 &  N/A\\
        \bottomrule
    \end{tabular}
\end{table}
\begin{table}[!h]
    \centering
    \caption{Convergence test for the TBS method. The units of the parameters are mentioned previously.}
    \label{tab:convergence_test}
    \renewcommand{\arraystretch}{1.6} 
    \begin{tabular}{l l l c }
        \toprule
        \textbf{Parameter} & \textbf{Constant parameters} &\textbf{Range} & \textbf{Convergence}  \\
        \midrule        
        $v_f/(fdx)$   &    & $\ge2$  & $\checkmark$ \\ 
        $\nu$   & $\alpha_0=5.5\times 10^{-10}$ & $\le 1.7$  & $\checkmark$ \\  
        $\tau$  & $\nu=1.5$, $\eta=-6.4\times 10^{-5}$        & $-10^{-5} ~ \mbox{to} ~ 10^{-6}$    & $\checkmark$ \\ 
        $\eta$          & $\nu=1.5$, $\tau=-4.3\times 10^{-8}$        &$ -10^{-3}~ \mbox{to} ~10^{-3}$    & $\checkmark$ \\ 
        $v_s$  & $a=1$, $v_f=1500$    &  1100-3750  & $\checkmark$ \\ 
        $B/A$    &      &no limit    & $\checkmark$ \\ 
        \bottomrule
    \end{tabular}
\end{table}
\section{Discussion and conclusions}
In this study, we have provided detailed derivations of the TBS framework for transducer field calculations for a series of realistic media, namely, homogeneous and lossless, absorbing and dispersive, inhomogeneous and nonlinear media. Accordingly, the source and potential terms have been identified for different cases by examining the conventional wave equations. After that the Green's function method has been applied to calculate the steady state solutions. The solutions are presented in terms of infinite series sums. In each case, field and phase characteristics for a line source are investigated. To compare the TBS results, k-Wave simulations have also been conducted. Perfect match between the field maps facilitated by the two methods can be seen. Phase maps also exhibit excellent match. The mean phase error between the TBS and k-Wave outcomes was found to be 1.5 milli-radians. The results suggest that the frequency domain and time domain approaches provide consistent results, with only minimal discrepancies that may arise due to numerical approximations, interpolation artifacts, or differences in boundary conditions. 

The numerical value of $\epsilon$ was chosen to be $\epsilon=0.8k_f^2$ in the present study \cite{Osnabrugge, Anuj_JMO, RKS1, RKS2, UM_Photoacoustics}. Note that the medium becomes more attenuating and thus, the disturbance spreads slowly as the strength of $\epsilon$ increases. As a result of that, the TBS technique takes more iterations to converge. In other words, number of iterations can be reduced by decreasing $\epsilon$ \cite{RKS1}. Further, the TBS method does not impose any restriction on the choice of magnitude of $\epsilon$, however,  $\epsilon \ge \mbox{max}(k_s^2-k_f^2)$ is required for the convergence of the CBS method for a heterogeneous medium \cite{Osnabrugge}. As mentioned earlier, CBS method works well even when large and strong scattering region/regions is/are present within the computational domain \cite{UM_Photoacoustics}. Attempts will be made in future to implement the CBS algorithm for transducer pressure field calculation in realistic media.
The memory requirement for the TBS formulation was about 2.7 GB for the linear media, respectively. The same for the k-Wave methodology was nearly 50 MB. The TBS technique does not take long iterations to converge as can be seen from Table \ref{tab:execution_time}. The number of iterations (column 2) has been counted to be approximately 135 for the fundamental frequency. The execution time for each case is also given in the same table (column 3) and it depends upon the number of matrix operations involved for calculating the field. For instance, computation time for absorbing and dispersive medium is higher than the lossless medium since Eq. (\ref{TBS_solution_absorbing-dispersive}) is more complex than Eq. (\ref{TBS_solution_lossless}). The average execution time is estimated to be around 7.5 s for $f=1007080$ Hz, whereas the same quantity for the k-Wave simulation is nearly 569 s. Therefore, nearly $76 \times$ speed up can be achieved by the TBS method. It may be pointed out that a parallel CPU code took 21 s for the homogeneous case in the same computer specifed in Table \ref{tab:HPC_specs} (speed up approximately $4 \times$). All details for the other CPU codes can be found in the same table (Table \ref{tab:execution_time}). It is also visible from Table \ref{tab:execution_time} that number of iterations vis-{\' a}-vis computation time increases progressively as the frequency increases for the nonlinear medium. Note that convergence reaches quickly if the initial field (0th iteration) is strong. Nevertheless, higher iteration is required if the initial field is weak (see for higher harmonics).

Table \ref{tab:convergence_test} highlights the impact of various parameters on the convergence of the TBS method. At first, it is observed that the TBS method converges until 2 points per wavelength is taken. For the second medium, convergence can be achieved for $\nu \le 1.7$ for the given $\alpha_0$, beyond which it diverges. Further, $\tau > -4.3 \times 10^{-5}$ and $\eta >-6.4\times 10^{-3} $ for convergence (see rows 4 and 5, respectively). For the fourth medium, the TBS method produces faithful results over a large range of sound-speed mismatch $v_s=1100$ to 3750 m/s. We have seen that convergence of the TBS technique is independent of the magnitude of $B/A$. This is obvious because $B/A$ or $\xi$ controls the strength of $S$ at higher harmonics, which may be arbitrary ($V$ is more critical than $S$ for convergence of the TBS algorithm). 

It is important to note that boundary condition plays a crucial role for obtaining accurate results in numerical field calculations. For the TBS method, we applied a Sigmoid function to dampen the outgoing wave. This function has been phenomenologically chosen and found to work exceedingly well for fundamental frequency as well as harmonics [see first row of Fig. \ref{Nonlinear}]. The function at first very slowly decays from unity and thus wave would not reflect back; on the other hand, it decays very slowly attaining zero, ensuring that the wave would not reemerge from the opposite boundary while leaving a boundary (minimizing wave wrapping). Wave wrapping is a serious issue that is associated with the extensive usage of the forward and inverse fast Fourier transforms. The k-Wave method uses a perfectly matched layer to attenuate the signal. It implements an anisotropic method. Essentially, the normal component (with respect to the boundary surface) of the particle velocity is gradually reduced leaving the parallel component unaltered. It may be pointed out that in case of CBS technique, a potential was assigned to each boundary layer causing absorption of the outgoing wave. The same group has also developed a method, which builds a pretty thin absorbing layer (much thinner than that of the present work) \cite{Vellekoop}.

The numerical study has been carried out for a 2D computational domain. Subsequently, the pressure field generated by a linear array of sources (line source) is computed for different realistic tissue properties. A practical diagnostic ultrasound array transducer is a 3D system (where small rectangular piezoelectric elements are glued on a backing layer in an ordered manner). We plan to implement the TBS/CBS method for realistic transducers (3D systems) in the near future. 
In conclusion, the TBS method is a standard and established technique to solve inhomogeneous Helmholtz equation. Conventionally, it has been extensively used to solve such an equation arising in potential scattering problems \cite{Sakurai},\cite{BransdenJoachain}. The TBS algorithm has been implemented in this work to yield solution to the same equation occurring in the context of biomedical ultrasonics. Note that acoustic wave propagation, through a homogeneous and lossless tissue medium, can essentially be modeled by the homogeneous Helmholtz equation when the region of interest does not contain any acoustic source. Nevertheless, the equation becomes inhomogeneous for a lossy or a nonlinear tissue medium or when a tissue retains scatterers (due to small fluctuations in physical and chemical properties) even in the absence of an acoustic source. Acoustic field distributions were computed for a line source mimicking a linear array transducer (in 2D) by the TBS method for various realistic media mentioned above; k-Wave simulations were also conducted for comparison. GPU-accelerated numerical computations were carried out. Perfect agreement have been noticed between the TBS and k-Wave results. The TBS method can provide a map of the pressure field within approximately 10 s for the computational domain considered in this study. The computation time for this approach is found to be 76 times (average) faster than the later. The TBS technique has emerged as a universal numerical method for accurate calculations of pressure field for all possible conditions associated with real tissue media.  
\section{Acknowledgments}
The computational results reported in this work were performed on the Central Computing Facility of IIITA, Prayagraj. UM thanks the members of the Biomedical Imaging Laboratory for their support during this work and UGC NFSC (\#F. 82-44/2020 (SA-III)) for providing the fellowship. RKS acknowledges the financial support received from SERB (\# CRG/2023/003278). BTC thanks EPSRC, UK, grants EP/W029324/1 and EP/T014369/1. 
%
%
%
%
%%%%%%%%%%%%%%%%%%%%%%%%%%%%%%%%%%%%%%%%%%%%%%%%%%%%%%%%%%%%%%%%%%%%%%%%%%%%%%%%%%%%%%%%%%

%TC:endignore

%%%%%%%%%%%%%%%%%%%%%%%%%%%%%%%%%%%%%%%%%%

%
\clearpage
\newpage
%\
\section*{\Large{Supplementary materials}}
\setcounter{figure}{0}
\setcounter{table}{0}
\renewcommand{\thetable}{S\arabic{table}}
\renewcommand{\thefigure}{S\arabic{figure}}
\begin{figure*}[!h]
\centerline{\includegraphics[width=1.0\textwidth]{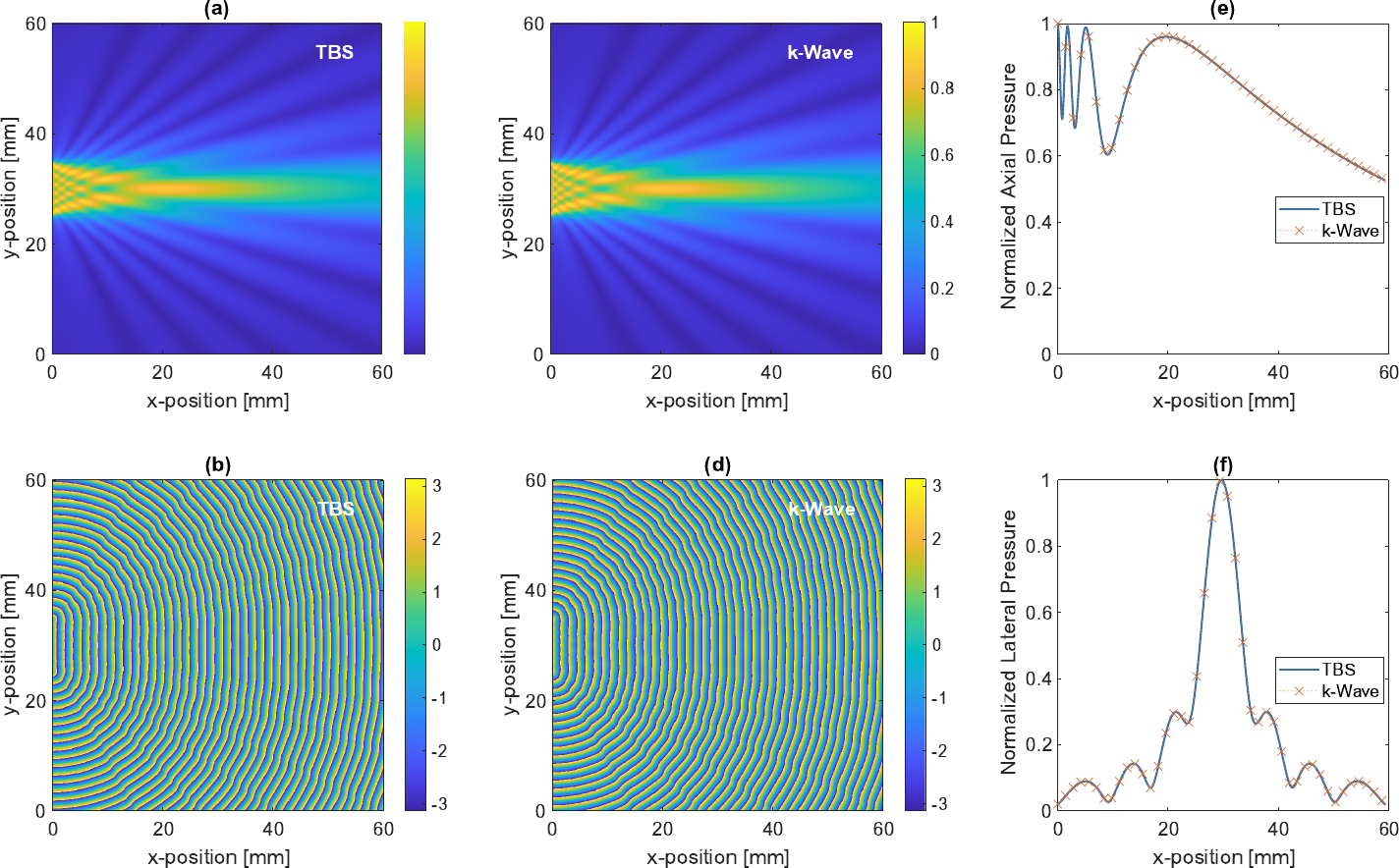}}
\caption{(a)-(b) Visualization of the steady state pressure field distribution (amplitude and phase) for the TBS scheme for an absorbing medium having $\alpha_0=5.5 \times 10^{-10}$ Np(rad/s)$^{\nu}$ m$^{-1}$ and $\nu=1.5$. (c)-(d) Same results for the k-Wave toolbox. On and off axes plots of pressure data are shown in (e) and (f), respectively. The lateral line is $x=41$ mm away from the line source.}
\label{Absorbing}
\end{figure*}
\begin{figure*}[]
\centerline{\includegraphics[width=1.0\textwidth]{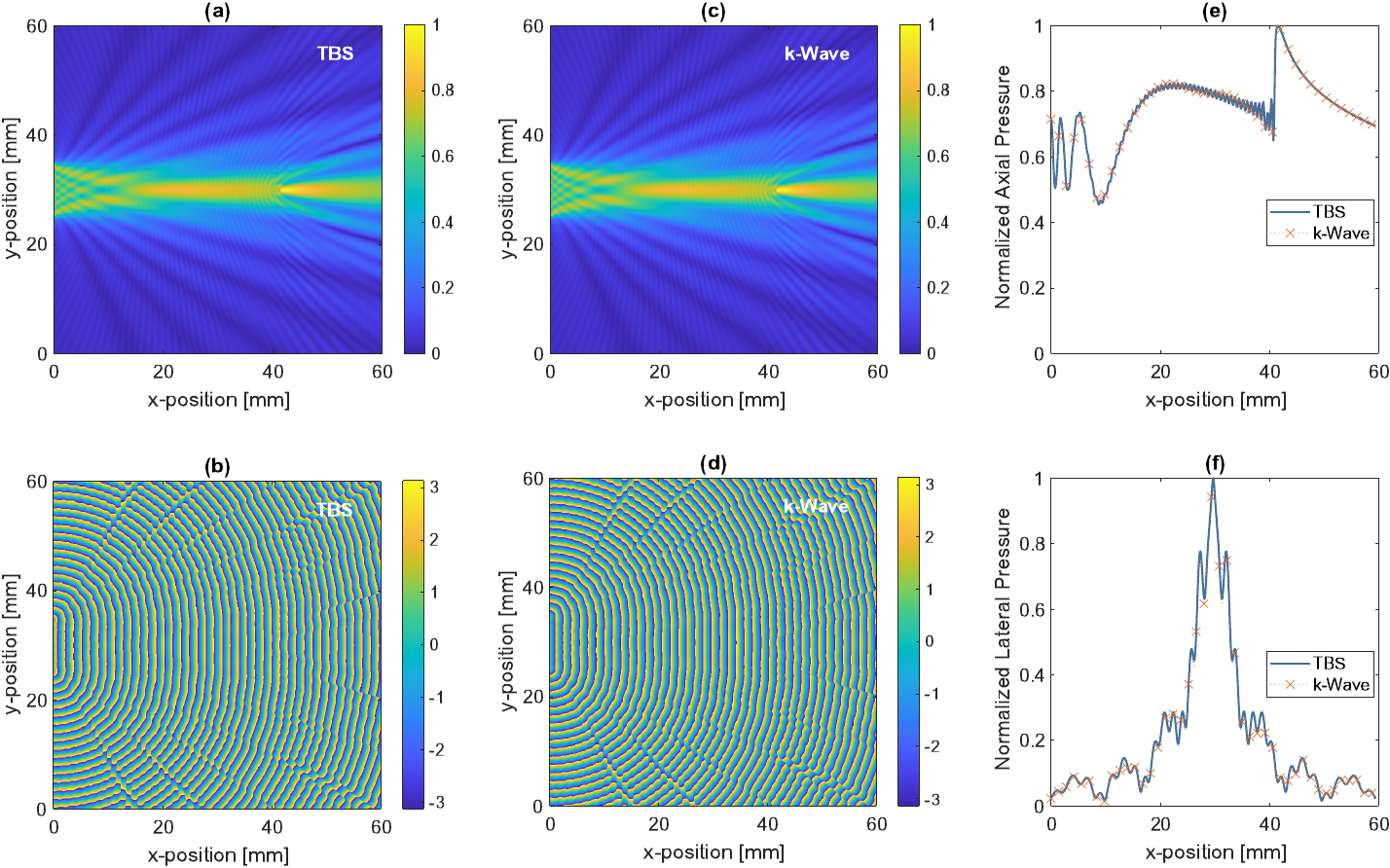}}
\caption{Same as \ref{Absorbing} but for an inhomogeneous medium. The propagating medium contains an inhomogeneity of $a=1$ mm with speed of sound $v_s = 1350$ m/s inside and $v_f = 1500$ m/s outside the scatterer.}
\label{Inhomogeneous1350}
\end{figure*}
\begin{figure*}[]
\centerline{\includegraphics[width=1.0\textwidth]{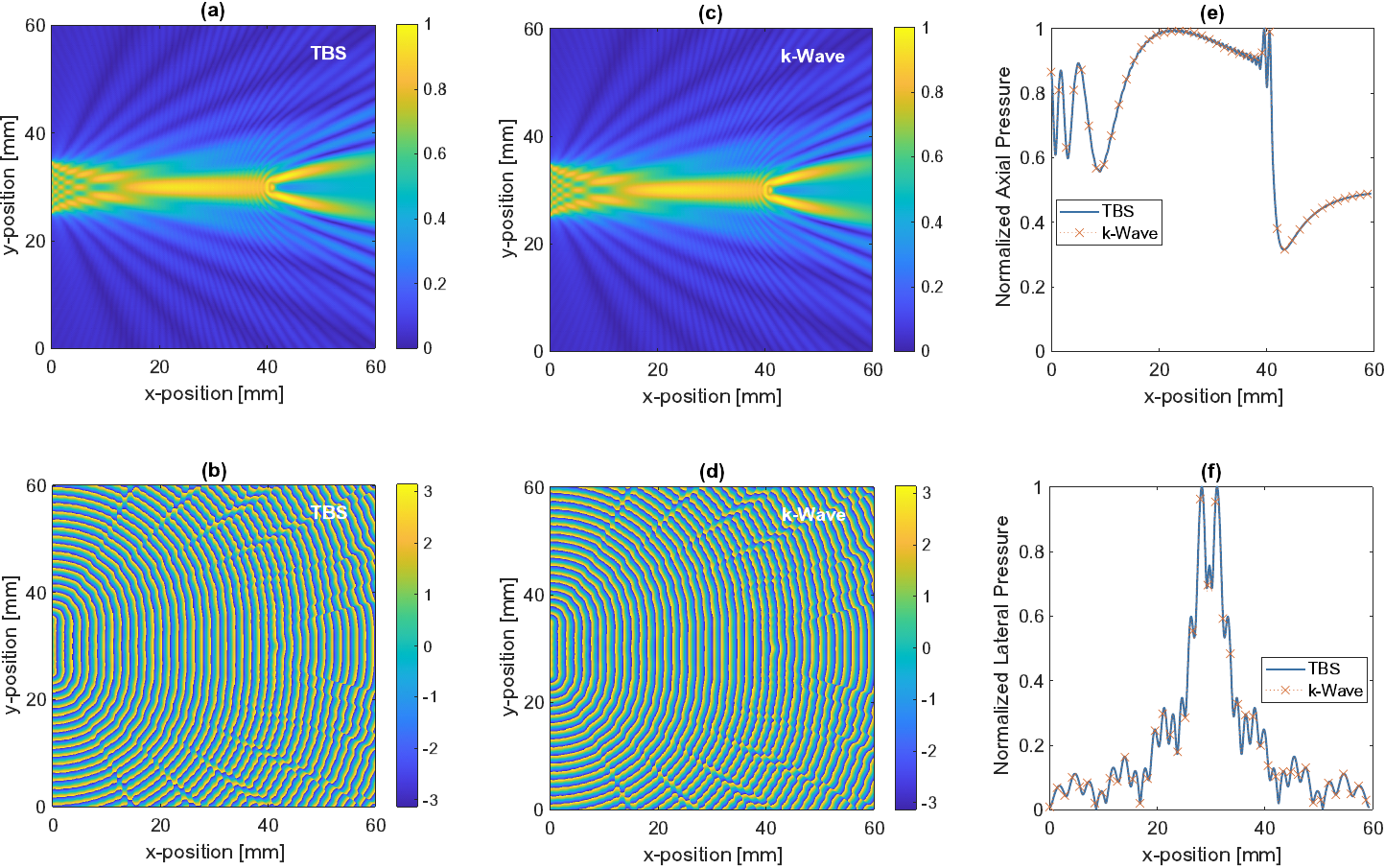}}
\caption{ Same as Fig. \ref{Inhomogeneous1350} but for  $v_s = 1800$ m/s}.
\label{Inhomogeneous1800}
\end{figure*}
\end{document}